\documentclass[prb,aps,twocolumn,amsmath,nofootinbib,superscriptaddress,10pt]{revtex4-2}
\pdfoutput=1
\usepackage[left=1.5cm,right=1.5cm,top=2.5cm,bottom=2.5cm]{geometry}

\usepackage[utf8]{inputenc}
\usepackage{natbib}
\usepackage{multirow}
\usepackage{physics}
\usepackage{natbib}
\usepackage{cancel}
\usepackage{comment}
\usepackage[english]{babel}
\usepackage[pdftex]{graphicx}
\usepackage[lflt]{floatflt}
\usepackage{epstopdf}
\usepackage{subcaption}
\usepackage{ragged2e}
\DeclareCaptionJustification{justified}{\justifying}
\usepackage[final]{hyperref} 
\hypersetup{
	colorlinks=true,       
	linkcolor=red,        
	citecolor=red,        
	filecolor=magenta,     
	urlcolor=blue         
}
\usepackage{chngcntr}

\usepackage[normalem]{ulem}

\numberwithin{equation}{section}
\numberwithin{figure}{section}
\numberwithin{table}{section}



\begin{document}

\title{Quantum fields in a cold atomic simulator: relaxation and phase locking in tunnel-coupled 1D bosonic quasi-condensates}

\author{Bence Fitos}\email{fitos5bence@gmail.com}
\affiliation{Department of Theoretical Physics, Institute of Physics, Budapest University of Technology and Economics, H-1111 Budapest, M{\H u}egyetem rkp.~3.}
\affiliation{HUN-REN-BME-BCE Quantum Technology Research Group, Institute of Physics, Budapest University of Technology and Economics, M{\H u}egyetem rkp.~3., H-1111 Budapest, Hungary}
\author{G\'abor Tak\'acs}\email{takacs.gabor@ttk.bme.hu}
\affiliation{Department of Theoretical Physics, Institute of Physics, Budapest University of Technology and Economics, H-1111 Budapest, M{\H u}egyetem rkp.~3.}
\affiliation{BME-MTA Momentum Statistical Field Theory Research Group, Institute of Physics, Budapest University of Technology and Economics, M{\H u}egyetem rkp.~3., H-1111 Budapest, Hungary}
\date{December 8, 2025}

\begin{abstract}
We consider a prime example of simulating interacting relativistic QFT with cold atoms: the realisation of the sine-Gordon model by tunnel-coupled quasi-1D Bose gases. While experiments have shown that it can realise the sine-Gordon model in equilibrium, studies of non-equilibrium dynamics have revealed phase-locking behaviour that contrasts with predictions from sine-Gordon field theory. Here, we examine a one-dimensional field-theoretic model of the system and find that the phase-locking behaviour can be understood in terms of the longitudinal harmonic trap, and that the additional degrees of freedom observed in the experiment do not appear to play a significant role. Therefore, the experimental setup provides a good simulator of the sine-Gordon quantum field theory, even out of equilibrium, if the inhomogeneous background induced by the trap is taken into account. Furthermore, our results support the idea that modifying the longitudinal trap to a box shape should result in agreement with standard sine-Gordon dynamics. The main remaining open issues are accounting for 3D corrections and modelling the effect of the boundaries.
\end{abstract}

\maketitle

\tableofcontents

\section{Introduction}

Feynman's idea of using quantum systems to simulate other quantum systems \cite{Feynman:1981tf}, particularly quantum field theories, has found a promising avenue in experiments with ultracold atoms \cite{2010PhRvL.105s0403C}, which leverage the high degree of control and tunability offered by ultracold atomic systems to mimic the behaviour of quantum particles and fields, and offer a pathway to explore phenomena beyond the reach of classical computation. Engineering and manipulating isolated quantum systems has now become routine in experiments with trapped ultracold atoms \cite{2006Natur.440..900K,2007Natur.449..324H,2012NatPh...8..325T,2012Sci...337.1318G,2013NatPh...9..640L,2013PhRvL.111e3003M,2013Natur.502...76F,2015Sci...348..207L}. In particular, they provide access to the out-of-equilibrium dynamics of quantum systems, which has been at the forefront of research in recent years \cite{2010NJPh...12e5006C,2011RvMP...83..863P,2016RPPh...79e6001G,2016JSMTE..06.4001C}. 

Here we consider the sine-Gordon model, which is a paradigmatic interacting integrable quantum field theory that describes the low-energy physics of numerous physical systems \cite{1999PhRvB..60.1038A, 2009PhRvB..79r4401U, 2004PhRvL..93b7201Z,2019PhRvB.100o5425R, 2021NuPhB.96815445R,2005odhm.bookE, Giamarchi:743140, 2010Natur.466..597H, 2000cond.mat.11439C}. An interesting feature of the model is its integrability, which provides exact results for numerous quantities, ranging from scattering amplitudes and form factors to expectation values of local observables \cite{Zamolodchikov:1978xm, 1992ASMP...14.....S, 1997NuPhB.493..571L}. 

The sine-Gordon model can be realised experimentally with two Josephson-coupled one-dimensional bosonic quasi-condensates\footnote{Alternative cold atom realisations can be found in Refs. \cite{2010PhRvL.105s0403C,2010Natur.466..597H,2017Natur.545..323S,2023PRXQ....4c0308W}; additional proposals include realisations via quantum circuits \cite{2021NuPhB.96815445R} or coupled spin chains \cite{2022PhRvB.106g5102W}. } \cite{2007PhRvB..75q4511G,2013NatPh...9..640L,2017Natur.545..323S}, whereas ultra-cold atoms are trapped in an elongated double-well potential, effectively limiting the physics to one spatial dimension. Bosonisation of the one-dimensional description of the many-body system \cite{2007PhRvB..75q4511G}, leads to a sine-Gordon model realised by the relative phase between the condensates, which is weakly coupled to a Luttinger liquid accounting for their common (average) phase. The latter coupling breaks integrability; however, equilibrium correlation functions agree well with sine-Gordon predictions \cite{2017Natur.545..323S,2018PhRvA..98b3613B}, indicating that in thermal equilibrium the coupled condensates can be regarded as a good quantum simulator of the sine-Gordon quantum field theory. 

Out-of-equilibrium dynamics can be investigated using the paradigmatic protocol of quantum quench \cite{2006PhRvL..96m6801C,2007JSMTE..06....8C}, which involves preparing the system in an equilibrium state and then suddenly changing some physical parameter, e.g., a coupling in a Hamiltonian. In the coupled condensates, a natural protocol involves splitting a single one-dimensional condensate, imprinting a relative phase between them, and then switching on the tunnel coupling. Theoretical computations based on pure sine-Gordon dynamics predict a rephasing \cite{2013PhRvL.110i0404D,2015PhRvA..91b3627F,2017EPJST.226.2763F,Horvath2019}, in which the relative phase of the condensates relaxes to a finite value. 
\begin{figure*}
    \centering
    \includegraphics{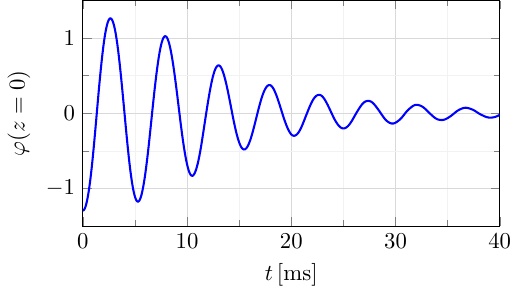}
    \includegraphics{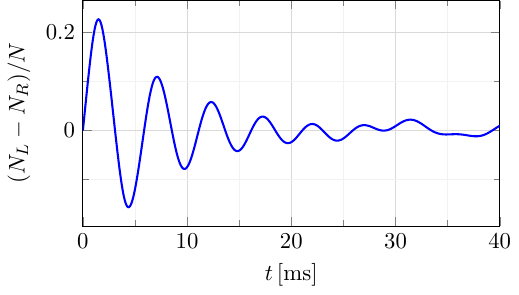}
    \includegraphics{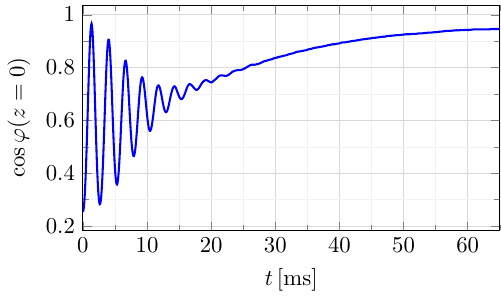}
    \includegraphics{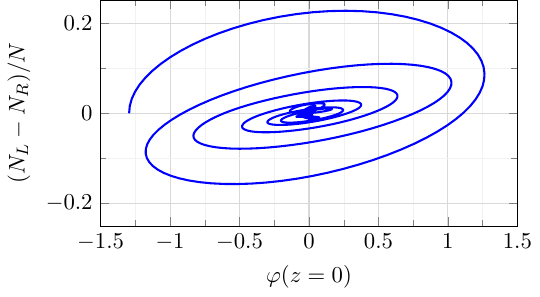}
    \caption{TWA results of the phase and the total particle number difference of a Josephson quench. We also illustrated the evolution of $\cos\varphi$ signalling phase locking with $\langle\cos\varphi\rangle_\text{eq} \approx 0.945$. The expectation value of $\varphi$ is taken in the middle of the condensate ($z = 0$). Parameters: $N = 3300,\, \omega_{||} = 2\pi\cdot 12\,\text{s}^{-1},\, J/(2\pi\hbar) = 19.0\,\text{Hz},\, T = 0, \sigma_N = 300,\, \varphi_0 = -1.3(2)$. 
    }
    \label{fig:comparism}
\end{figure*}
In contrast, the experiment found a rapid phase locking \cite{2018PhRvL.120q3601P} between the condensates, which can be parameterised in terms of a two-mode model extended with a dissipative term \cite{1999PhRvA..60..487M,2018PhRvA..98f3632P}. The main experimental findings can be captured using a stochastic Gross-Pitaevskii description \cite{2021PhRvR...3b3197M}, which, however, steps outside the framework of the quantum field theoretic (sine-Gordon) description. The alternative approach, involving a self-consistent Hartree-Fock approximation \cite{2019JSMTE..08.4012V,2020ScPP....9...25V,2021ScPP...10...90V}, is not applicable in the parameter regime relevant to the experiment, which involves a significantly higher number of particles than those used in the model calculation.

In this paper, we consider the dynamics of the system in the effective coupled sine-Gordon-Luttinger field theory description, using the so-called truncated Wigner approximation \cite{1998PhRvA..58.4824S,2001PhRvL..87u0404S,2010PhRvA..81f3605B}. In fact, this has already been applied to the sine-Gordon model, resulting in universal rephasing dynamics \cite{2013PhRvL.110i0404D}. The validity of the TWA method can be confirmed by using Hamiltonian truncation \cite{2019PhRvA.100a3613H,2024PhRvB.109a4308S}, which, however, cannot reach the experimentally relevant parameter regime. Here we take into account the longitudinal trapping potential, as well as the coupling between the sine-Gordon and Luttinger degrees of freedom and demonstrate that this approach captures the phase locking dynamics as illustrated in Fig. \ref{fig:comparism}. For qualitative comparison with the experiment, the parameters are selected according to Ref.~\cite{2018PhRvL.120q3601P}.

The outline of the paper is as follows. Section \ref{sec:model_and_method} reviews the derivation of the effective field theoretical description for the coupled condensates and the implementation of the truncated Wigner approximation. In Section \ref{sec:breaking_down_the_model}, we consider three crucial ingredients of the model: the nonlinearity inherent in the sine-Gordon self-interaction, the spatial inhomogeneity resulting from the trapping potential, and the coupling between the relative (sine-Gordon) and common (Luttinger) modes. Section \ref{sec:fulldyn} presents the results for the full dynamics in the experimental parameter regime. In section \ref{sec:comparison}, we analyse the scaling of the Josephson frequency and relaxation time of the relative phase as a function of various input parameters of the experiment and compare the findings to previous results. In addition, we compare our results to the leading order behaviour of small quenches described by a free massive boson approximation of the sine-Gordon field. Finally, we present our conclusions in Section \ref{sec:conclusions}. We also added a summary of useful relations between different parameters in Appendix \ref{app:parameters}.

\section{Model and methodology}\label{sec:model_and_method}

\subsection{The effective 1D description}

In this section, we describe the quasi-one-dimensional model of a single elongated bosonic quasi-condensate. We start from the grand canonical Hamiltonian of a three-dimensional interacting Bose gas:
\begin{widetext}
\begin{align}
    H = & \int d^3\vec{r}\Psi^\dagger(\vec{r}) \left(-\frac{\hbar^2}{2m}\nabla^2 + V(\vec{r}) - \mu\right)\Psi(\vec{r}) + \int d^3\vec{r}\,d^3\vec{r}'\Psi^\dagger(\vec{\vec{r}})\Psi^\dagger(\vec{r}')U_\mathrm{eff}(\vec{r}-\vec{r}')\Psi(\vec{r}')\Psi(\vec{r})\,,\nonumber \\    &\left[\Psi(\vec{r}),\Psi^\dagger(\vec{r}')\right]=\delta(\vec{r}-\vec{r}')
\end{align}
\end{widetext}
where $m$ is the atomic mass, $V(\vec{r})$ is the 3D trapping potential and $\mu$ is the chemical potential. For low temperatures, the effective interaction $U_\text{eff}$ can be approximated by a delta potential 
\begin{equation}
    U_\mathrm{eff}(\vec{r}-\vec{r}')=\frac{4\pi\hbar^2}{m}a_s \delta(\vec{r}-\vec{r}')
\end{equation}
characterised by the $s$-wave scattering length $a_s$.

Switching on a radial trapping potential, e.g. of the form
\begin{equation}
    \frac{1}{2}m \omega_{\perp}^2 (x^2+y^2) 
\end{equation}
in the $xy$ plane leads to an elongated condensate, which can be modelled as an effectively one-dimensional Lieb-Liniger gas
\begin{widetext}
\begin{align}
    H = \int dz\, \psi^\dagger(z) \left(-\frac{\hbar^2}{2m}\partial_z^2 + V(z) - \mu + \frac{g_\mathrm{1D}}{2}\psi^\dagger(z)\psi(z)\right)\psi(z)\,,
\end{align}    
\end{widetext}where the effective one-dimensional coupling can be expressed in terms of the $s$-wave scattering length and the frequency $\omega_\perp$ of the transversal trapping potential \cite{1998PhRvL..81..938O}
\begin{equation}
g_\mathrm{1D}\approx\frac{2\hbar^{2}a_{s}}{ml_{\perp}^{2}}\left(1-1.036\frac{a_{s}}{l_{\perp}}\right)^{-1}\;.
\end{equation}
Here $l_{\perp}=\sqrt{\hbar/(m\omega_{\perp})}$ is the transverse oscillator
length associated with the frequency $\omega_{\perp}$ of the radial confining
potential, and $a_{s}$ the three-dimensional $s$-wave scattering
length of the atoms. In the weak interaction regime $a_{s}\ll l_{\perp}$ the result simplifies $g_\mathrm{1D}\approx2\hbar\omega_{\perp}a_{s}.$

We assume that the system is in the quasi-condensate regime where the density of the condensate can be considered a classical function $\rho_0(z)$ dressed up by relatively smaller quantum fluctuations. To obtain a low-energy effective theory, it is useful to introduce the Madelung representation of the field
\begin{align} \label{eq:mad}
    \psi(z) = e^{i\varphi(z)}\sqrt{\rho_0(z) + \delta\rho(z)}\,,
\end{align}
where $\hat\varphi(z)$ and $\delta\hat\rho(z)$ are real-valued fields. The classical background density profile $\rho_0(z)$ satisfies the Gross--Pitaevskii equation
\begin{align}
    \left(-\frac{\hbar^2}{2m}\partial_z^2 + V(z) -\mu + g_\mathrm{1D}\rho_0(z)\right)\sqrt{\rho_0(z)} = 0\,,
    \label{eq:GP}
\end{align}
where the chemical potential $\mu$ is fixed by the total number of particles
\begin{align}
    N = \int dz \rho_0(z)\,.
\end{align}
In the experimental setup, the condensate is further confined in the $z$ direction by the quadratic potential
\begin{align}
    V(z) &= \frac{1}{2} m\omega_{||}^2z^2\,,
\end{align}
where $\omega_{||}\ll \omega_{\perp}$. Neglecting the kinetic term in \eqref{eq:GP} (local density approximation) gives the Thomas--Fermi density profile
\begin{align}
    \rho_0(z) &= n_0\left(1 - \frac{z^2}{R_\text{TF}^2}\right)\,,
    \label{eq:TF_profile}
\end{align}
where
\begin{align}
    R_\mathrm{TF} = \frac{L}{2}\,,\quad
    n_0 = \frac{3N}{2L}\,,
\end{align}
with $L$ giving the full length in the $z$ direction. 

To the leading order approximation in the density fluctuations $\delta\hat\rho$, the system is then described by an inhomogeneous Luttinger--liquid Hamiltonian \cite{2004cond.mat..9230S}
\begin{align}
    H_\mathrm{TLL} &= \frac{\hbar}{2\pi}\int dz\left(\nu_N(z)\left(\pi\delta\rho\right)^2 + \nu_J(z)\left(\partial_z\varphi\right)^2\right)\,
\label{eq:TLL}\end{align}
where the density and phase stiffness are given by \cite{2004JPhB...37S...1C,2011RvMP...83.1405C}
\begin{align}
    \begin{aligned}
    \nu_N(z) &=\frac{1}{\pi\hbar}\left.\frac{\partial\mu}{\partial\rho_0}\right|_{\rho_0=\rho_0(z)}\approx \frac{g_\mathrm{1D}}{\pi\hbar}\,\\
    \nu_J(z) &= \frac{\pi\hbar\rho_0(z)}{m}\,.
    \end{aligned}
\end{align}
where the first approximation holds in the weakly interacting regime
\begin{equation}
    \gamma=\frac{mg_\mathrm{1D}}{\hbar^2n_\mathrm{1D}}\ll 1\,.
\end{equation}
The Hamiltonian can be written in the form
\begin{align}
    H_\mathrm{TLL} &= \frac{\hbar c_0}{2}\int dz\left( \frac{\pi}{K_0} \delta\rho^2 + f(z)\frac{K_0}{\pi} (\partial_z\varphi)^2\right)\,,
\label{eq:dimlessTLL}\end{align}
where 
\begin{equation}
    n_0 = \rho_0(z = 0)
\end{equation}
is the density, 
\begin{align}
    f(z) = \frac{\rho_0(z)}{n_0}
\end{align}
is the dimensionless density profile,
\begin{align}
    c_0 = \sqrt\frac{g_\mathrm{1D}n_0}{m}
\end{align}
is the sound velocity, and
\begin{align}
    K_0 = \pi\hbar \sqrt\frac{n_0}{mg_\mathrm{1D}}
\end{align}
is the (dimensionless) Luttinger parameter, and the index $0$ indicates that all of them are specified by their local values at the midpoint $z=0$.

The spectrum of the quadratic Hamiltonian \eqref{eq:dimlessTLL} can be obtained in terms of decoupled bosonic modes satisfying the classical equation of motion in terms of the renormalised field $\phi(z) = \sqrt{K_0 / \pi} \,\varphi(z)$
\begin{align}
    -\partial_t^2\phi + \partial_z\left(f\partial_z\phi\right) &= 0\,,
\end{align}
for which the solutions can be written in the form
\begin{align}
    \phi_{n\pm}(t,z) &= e^{\pm i\omega_n t}Z_n(z)\,,
\end{align}
where
\begin{align}\label{Eq:Zn}
    \partial_z\left(f\partial_z Z_n\right) &= -\omega_n^2 Z_n\,,
\end{align}
{with the boundary conditions}
\begin{align}
    \lim_{z\rightarrow\pm L/2}f(z) Z_n(z) = 0\,,
\end{align}
{obtained from the Hamiltonian \eqref{eq:dimlessTLL},} where we choose all the $Z_n$'s as real-valued functions, which form an orthonormal basis
\begin{align}
    \int_{L/2}^{L/2}dz Z_n(z) Z_m(z) &= \delta_{nm}\,.
\end{align}
Then $\phi$ can be decomposed into eigenmodes
\begin{align}
    \phi(t,z) &= \sum_n\frac{Z_n(z)}{\sqrt{2\omega_n}}\left(a_ne^{-i\omega_n t} + a_n^\dagger e^{-i\omega_n t}\right)\,,
\end{align}
where $a_n$ and $a_n^\dagger$ are bosonic creation and annihilation operators, from which the Hamiltonian becomes
\begin{align}
    H_\mathrm{TLL} &= \hbar c_0 \sum_n \omega_n a_n^\dagger a_n + \mathrm{const}\,.
\end{align}
In the case of the Thomas-Fermi density profile \eqref{eq:TF_profile} we have $f(z) = 1 - 4z^2/L^2$ and the $Z_n$ functions can be expressed by Legendre polynomials:
\begin{align}
    Z_n(z) &= \sqrt{\frac{2n+1}{L}} P_n\left(\frac{z}{L/2}\right)\,,
\end{align}
with the corresponding frequencies given by
\begin{align}
    \omega_n &= \frac{1}{L}\sqrt{(2n+1)^2-1}\,.
\end{align}

The decomposition of the field into eigenmodes can be written as
\begin{align}
    \begin{aligned}
    \tilde\phi_n &= \int dz Z_n(z)\phi(z)\,,\\
    \tilde\Pi_n &= \int dz Z_n(z) \Pi(z)\,,
    \end{aligned}
\label{eq:eigenmode_decomp}\end{align}
with the inverse transformations
\begin{align}
    \begin{aligned}\label{eq:invtr}
    \phi(z) &= \sum_n Z_n(z)\tilde\phi_n\,,\\
    \Pi(z) &= \sum_n Z_n(z)\tilde\Pi_n\,.
    \end{aligned}
\end{align}
{Note that the treatment of the boundary conditions is rather subtle. Near the edges of the condensate, the density approaches zero, and the Thomas-Fermi approximation ceases to be valid. For typical experimental parameters, however, the particle density remains sufficiently large over most of the condensate, such that the bulk Hamiltonian \eqref{eq:dimlessTLL} provides a valid description everywhere except within a narrow region near the boundaries. For simplicity, we do not attempt to model these edge regions explicitly; instead, we extend the bulk Hamiltonian to the entire condensate. For local observables measured near the centre of the condensate, the effects of this approximation can only become relevant once signals originating from the boundaries have had sufficient time to propagate to the centre.}

\subsection{The tunnel coupling}

The experimental setup corresponds to coupling two quasi-condensates $\psi_{1,2}(z)$ via a tunnelling Hamiltonian
\begin{align}
    H_I = -J\int dz \left(\psi_1^\dagger(z) \psi_2(z) + \psi_2^\dagger(z) \psi_1(z)\right)\,.
\end{align}
Using the Madelung representation Eq.~\eqref{eq:mad} for of $\psi_{1,2}$ shows that the coupling between the two condensates depends on the relative phase $\varphi_1(z) - \varphi_2(z)$. Therefore, we introduce the \emph{relative} and \emph{common} degrees of freedom
\begin{align}
    \begin{aligned}
    \varphi^r(z) := \varphi_1(z) - \varphi_2(z)\,,\\
    \varphi^c(z) := \frac{\varphi_1(z) + \varphi_2(z)}{2}\,,\\
    \delta\rho^r(z) := \frac{\delta\rho_1(z) - \delta\rho_2(z)}{2}\,,\\
    \delta\rho^c(z) := \delta\rho_1(z) + \delta\rho_2(z)\,.
    \end{aligned}
\end{align}
To the lowest order in density fluctuation, the relative mode is described by the sine-Gordon model with the interaction Hamiltonian
\begin{align} \label{Eq:cc}
    H^{(0)}_I &= - 2J \int dz\,\rho_0(z) \cos( \varphi^r(z))\,,
\end{align}
while the common mode remains an independent Luttinger liquid. Higher-order corrections couple the relative to the common mode, and the next-to-leading order gives the term
\begin{align} \label{eq:cmc}
    H^{(1)}_I &= -J\int dz\, \delta \rho^c(z)\left(\cos( \varphi^r(z)) - 1\right)\,.
\end{align}
{In what follows, we retain this first subleading term and also examine its effect on the dynamics. Note that, while this correction remains regular at the boundaries, higher-order terms in the expansion diverge near the edges, despite being negligible over most of the condensate. To preserve a simple and self-consistent description, we therefore neglect these higher-order corrections.}

\subsection{Lattice discretisation}

The inhomogeneous Luttinger Hamiltonian \eqref{eq:TLL} can be discretised by a lattice $z_j=ja$ with lattice spacing $a$, and its dimensionless version (corresponding to condensate $s = 1,2$) can be written in the form
\begin{align} \label{Eq:discrTLL}
    &\tilde H^{\#}_{\mathrm{TLL},s} = 
    \nonumber\\
    &
    \frac12\sum_j \bigg(\Pi_{s,j}^2 
    + \frac{f(z_j) + f(z_{j+1})}{2}\left(\phi_{s,j+1} - \phi_{s,j}\right)^2\bigg)\,,
\end{align}
where
\begin{align}
    \begin{aligned}
    \tilde H^\# &= \frac{a}{\hbar c_0} H^\#\,,\\
    \Pi_{s,j} &= \sqrt\frac\pi{K_0} a\, \delta\rho_s(z_j)\,,\quad
    \phi_{s,j} = \sqrt\frac{K_0}\pi \varphi_s(z_j)\,.
    \end{aligned}
\end{align}

To characterise the tunnel coupling interaction suitably, we switch to relative and common degrees of freedom
\begin{align}
    \begin{aligned}
        \phi^r_j &= \frac{1}{\beta_r}\varphi^r(z_j)\,,\quad\Pi^r_j = a\beta_r\delta\rho^r(z_j)\,,\\
        \phi^c_j &= \frac{1}{\beta_c}\varphi^c(z_j)\,,\quad\Pi^c = a \beta_c\delta\rho^c(z_j)\,,
    \end{aligned}
\end{align}
where
\begin{align}
    \begin{aligned}
    \beta_r &= \sqrt\frac{2\pi}{K_0}\,,\quad
    \beta_c &= \sqrt\frac{\pi}{2 K_0}\,.
    \end{aligned}
\end{align}
The Luttinger Hamiltonian in terms of the relative and common degrees of freedom takes the same form as Eq.~\eqref{Eq:discrTLL}, while the dimensionless lattice regularised interaction terms corresponding to \eqref{Eq:cc} and \eqref{eq:cmc} can be expressed as
\begin{align}
    \begin{aligned}
        \tilde H_I^{(0)\#} &= \frac{2g L n_0}{N_s} \sum_j f(z_j)\cos(\beta_r\phi^r_j)\,,\\
        \tilde H_I^{(1)\#} &= \frac{g}{N_s} \sum_j \frac{\Pi^c_j}{\beta_c}\left(\cos(\beta_r \phi^r_j) - 1\right)\,,
    \end{aligned}
\end{align}
where $N_s = L/a$ is the number of lattice sites, and
\begin{align}
    \begin{aligned}
    g &= 2\pi \frac{J}{2\pi\hbar}\frac{L}{c_0}\,.
    \end{aligned}
\end{align}
Combining the free and interacting parts, the complete lattice regularised dimensionless Hamiltonian takes the form
\begin{widetext}
\begin{align}
    \begin{aligned} \label{eq:H}
        \tilde H^\# &= \tilde H_\mathrm{TLL,1}^\# + \tilde H_\mathrm{TLL,2}^\# + \tilde H_I^{(0)\#} + \tilde H_I^{(1)\#} 
        \\
        =\sum_j\biggl\{&\frac12\left[\left(\Pi_j^r\right)^2 + \frac{f(z_j) + f(z_{j+1})}{2}\left(\phi^r_{j+1} - \phi^r_j\right)^2\right]
        + \frac12\left[\left(\Pi_j^c\right)^2 + \frac{f(z_j) + f(z_{j+1})}{2}\left(\phi^c_{j+1} - \phi^c_j\right)^2\right]
        \\ 
        &- \frac{2g L n_0}{N_s^2} f(z_j)\cos(\beta_r\phi^r_j)
        - \frac{g}{N_s} \frac{\Pi^c_j}{\beta_c}\left[\cos(\beta_r \phi^r_j) - 1\right]\biggr\}\,.
    \end{aligned}
\end{align}
\end{widetext}

\subsection{The truncated Wigner approximation}

To simulate the quasi-condensate in the experimentally relevant parameter regime, we adopt a powerful semiclassical method, the truncated Wigner approximation (TWA) \cite{2003PhRvA..68e3604P, 2010AnPhy.325.1790P}, which is particularly suitable for calculating out-of-equilibrium expectation values. The TWA is based on the one-to-one mapping between quantum operators and ordinary functions defined in the (classical) phase space: for an arbitrary operator $\hat\Omega$ we define the \emph{Weyl symbol}
\begin{align}
    \Omega_W(\textbf{x},\textbf{p}) = \int d\textbf{x}' \left\langle \textbf{x} - \frac{\textbf{x}'}{2}\right|\hat{\Omega}\left|\textbf{x} + \frac{\textbf{x}'}{2}\right\rangle e^{i\textbf{p}\textbf{x}'}\,,
\end{align}
where $|\textbf{x}\rangle$ are the eigenstates of the position operator $\hat{\textbf{x}}$. In particular, the Weyl symbol of the density matrix with a suitable normalisation is called the Wigner function
\begin{align}
    W(\textbf{x},\textbf{p}) = \frac{1}{(2\pi\hbar)^{N_s}}\int d\textbf{x}' \left\langle \textbf{x} - \frac{\textbf{x}'}{2}\right|\hat{\rho}\left|\textbf{x} + \frac{\textbf{x}'}{2}\right\rangle e^{i\textbf{p}\textbf{x}'}\,,
\end{align}
where $N_s$ is the dimension of the $\textbf{x}$ and $\textbf{p}$ vectors.

The trace of a product of two operators can be expressed as a phase space integral of their Weyl symbols. In particular, for an expectation value, we get
\begin{align}
    \left\langle\hat\Omega\right\rangle = \int d\textbf{x}\,d\textbf{p}\,W(\textbf{x},\textbf{p}) \Omega_W(\textbf{x},\textbf{p})\,.
\end{align}
To incorporate time evolution, we approximate the trajectories of the phase space points with the classical paths
\begin{align}
    \left\langle\hat\Omega\right\rangle_\text{TWA}(t) = \int d\textbf{x}_0\,d\textbf{p}_0\,W(\textbf{x}_0,\textbf{p}_0) \Omega_W(\textbf{x}(t),\textbf{p}(t))\,,
\end{align}
where $\textbf{x}(t)$ and $\textbf{p}(t)$ are the solutions of the classical equations of motion with the initial conditions $\textbf{x}(0) = \textbf{x}_0$ and $\textbf{p}(0) = \textbf{p}_0$, respectively. The truncated Wigner approximation can be systematically derived as the leading order of a systematic expansion of the Keldysh path integral for quantum fields \cite{2003PhRvA..68e3604P,2010AnPhy.325.1790P}. Here, we follow the exposition in \cite{2019PhRvA.100a3613H}, to which we refer for further details on the implementation.

In the context of the coupled condensate system, the canonically conjugate classical variables are $\{\phi^r_j, \Pi^r_j\}_{j=1,\dots,N_s}$ and $\{\phi^c_j, \Pi^c_j\}_{j=1,\dots,N_s}$. The pre-quench state is the uncoupled condensate, $J = 0$, corresponding to the dimensionless Hamiltonian $\tilde H^\#_\mathrm{pre} = \tilde H^\#_{\mathrm{TLL},1}+\tilde H^\#_{\mathrm{TLL},2}$. The pre-quench Hamiltonian is quadratic and can be diagonalized with the linear transformation $\{(\phi^\alpha_j, \Pi^\alpha_j)\} \rightarrow \{(\tilde\phi^\alpha_n, \tilde\Pi^\alpha_n)\}, \alpha = r,c$ described in Eq.~\eqref{eq:eigenmode_decomp}.\footnote{We define an additional normalization factor $\sqrt a$ to eliminate the dimension of the $Z_n$ functions.} Note that, for a lattice regularised theory, the eigenmode decomposition \eqref{eq:eigenmode_decomp} transformation is established by an orthogonal matrix, which is the set of eigenvectors of the lattice regularisation of the differential operator in Eq.~\eqref{Eq:Zn}.

The ground state Wigner function is a Gaussian in each transformed mode $\{(\tilde\phi^\alpha_n, \tilde\Pi^\alpha_n)\}$, which allows for Monte-Carlo sampling. For $n \neq 0$, the standard deviations of the $n$th mode $(\tilde\phi^\alpha_n, \tilde\Pi^\alpha_n)$ is given by
\begin{align}
    \begin{aligned}
    \sigma_{\tilde\phi_n^\alpha} &= \frac{1}{\sqrt{2 a \omega_n}}\,,\\
    \sigma_{\tilde\Pi_n^\alpha} &= \sqrt\frac{a\omega_n}{2}\,.
    \end{aligned}
\end{align}
The zero mode energy is zero ($\omega_0 = 0$), therefore the ground state Wigner function is a uniform distribution on $[0,2\pi\sqrt{N_s}/\beta_\alpha)$ in variable $\tilde \phi_0^\alpha$ and a $\delta$ function (centred at $0$) in variable $\tilde\Pi_0^\alpha$.

The experimental quenching protocol sets the total relative phase
\begin{align}
    \varphi_0^r = \frac1L\int dz\,\varphi^r(z) = \frac{\beta_r}{\sqrt{N_s}}\tilde\phi_0^r
\end{align}
with some standard deviation $\sigma_{\varphi_0}$. Therefore, instead of a uniform distribution, we incorporate the experimentally given zero mode of the relative phase $\tilde\phi^r_0$ into the Wigner function: a normal distribution centred at $\sqrt{N_s}\varphi_0^r/\beta_r$ with standard deviation
\begin{align}
    \sigma_{\tilde\phi_0^r} &= \frac{\sqrt{N_s}}{\beta_r}\sigma_{\varphi_0}\,.
\end{align}
The corresponding conjugate variable $\tilde\Pi_0^r$ is chosen to be a Gaussian centred around $0$ with standard deviation minimising the Heisenberg inequality:
\begin{align}
    \sigma_{\tilde\Pi_0^r} &= \frac{\beta_r}{2\sqrt{N_s}\sigma_{\varphi_0}}\,.
\end{align}

The TWA simulation consists of taking a Monte-Carlo sample of initial values from which the real space functions are obtained via the inverse transformation Eq.~\eqref{eq:invtr}. The TWA simulation then consists of computing the classical time evolution from these initial conditions using the interacting Hamiltonian $J\neq 0$. Observables are computed by averaging the values obtained from individual trajectories. The Monte-Carlo error of these averages is estimated from the standard deviation of the data at each time point. The TWA has several significant advantages, including numerical stability and broad scalability, which enable negligible MC error even with limited computational resources. We note that TWA also allows for computations at finite temperature $k_BT = 1/\beta$, where each (non-zero) mode in the Wigner function remains Gaussian with a broadened standard deviation
\begin{align}
    \sigma^{(\beta)} &= \sigma\cdot\sqrt{\frac{e^{\beta\omega}+1}{e^{\beta\omega}-1}}\,.
\end{align}

\section{Breaking down the model}\label{sec:breaking_down_the_model}

\begin{figure*}[t]
    \centering
    \includegraphics{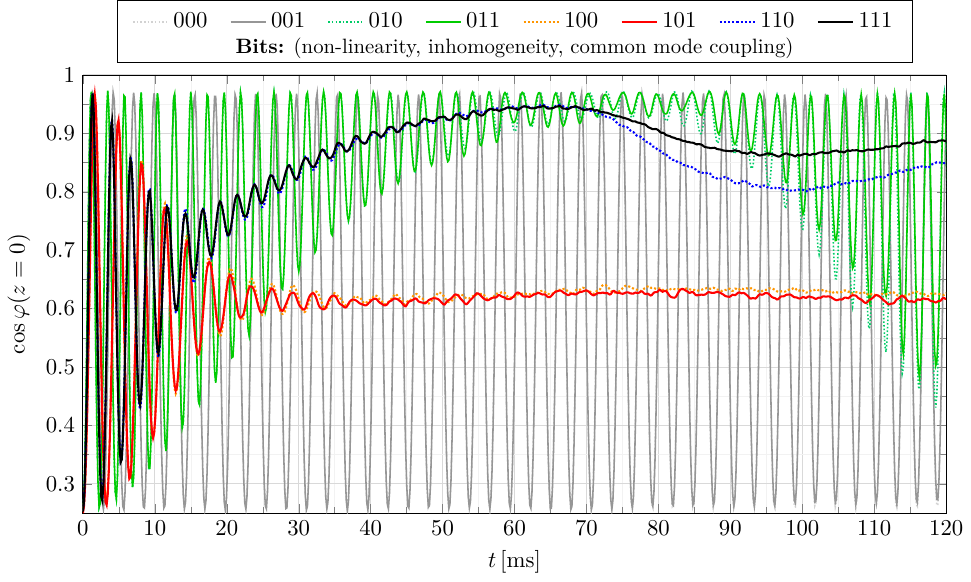}
    \caption{Eight scenarios compared. The binary codes indicate the presence or absence of the model's three characteristics: nonlinearity, inhomogeneity, and common-mode coupling. Parameters: $N = 3300, J/(2\pi\hbar) = 19.0\,\text{Hz}, T = 0.0\,\text{nK}, \varphi_0^r = -1.3, \sigma_{\varphi_0^r} = 0.2, \sigma_N = 0$.}
    \label{fig:8fold}
\end{figure*}

\subsection{The main scenarios}
To achieve a systematic understanding of the coupled condensate system, we dissect the model along its primary characteristics. We begin by examining the linearised sine-Gordon model on a homogeneous background. The full model is then constructed incrementally, with each step introducing a new aspect. For each of these refinements, we analyse its specific impact on the system's dynamics.

The subsequent analysis focuses on three crucial characteristics of the model:
\begin{itemize}
    \item \textbf{Non-linearity:} This concerns the validity of the quadratic approximation for the cosine potential in accurately describing the system's behaviour.
    \item \textbf{Inhomogeneity:} We examine the impact of the quadratic longitudinal trapping potential (resulting in an approximate Thomas-Fermi density profile) on the dynamics, contrasting it with a simple flat-bottom potential.
    \item \textbf{Common-mode coupling:} We examine the influence of the subleading interaction term, Eq.~\eqref{eq:cmc}, which couples the common and relative phase modes.
\end{itemize}
For a systematic comparison of these effects, we explore all $2^3=8$ combinations obtained by activating them. Throughout the different settings, we keep the local physical parameters constant at the midpoint $z=0$. Figure~\ref{fig:8fold} presents the resulting dynamics for the relative phase defined at the midpoint 
\begin{equation}
 \langle\cos\varphi\rangle(t)=\langle\cos\varphi^r(z=0,t)\rangle\,,
\end{equation} obtained using experimentally relevant parameters. Notably, the eight different scenarios lead to qualitatively distinct dynamic behaviours.

Our main finding is that the phase relaxation $\langle\cos\varphi\rangle(t) \rightarrow \approx 1$ can be attributed to the inhomogeneity resulting from the presence of the quadratic longitudinal trapping potential. Regardless of the other two settings, in the presence of an inhomogeneous background, the relative phase $\langle\cos\varphi\rangle(t)$ system approaches $\approx 1$ on the timescale of $2\pi/\omega_{||}$ determined by the longitudinal trap. However, once $\langle\cos\varphi\rangle(t)$ reached its peak, revivals appear, which decrease its value. Nevertheless, there is a well-defined plateau before the appearance of revivals when the $\langle\cos\varphi\rangle(t)$ is stabilised around a value close to $1$, indicating phase locking. In this regard, we note that even the expectation value of $\langle\cos\varphi\rangle$ does not exactly equal $1$ even in the (zero temperature) sine-Gordon vacuum \cite{1997NuPhB.493..571L} due to quantum fluctuations of the phase field $\varphi$. Even more deviation from $1$ is expected after a quantum quench, which injects finite energy density into the system. Therefore, even in phase locking, we expect residual fluctuations in the relative phase, leading to $\langle\cos\varphi\rangle(t)$ relaxing to a value slightly less than $1$.

The nonlinearity of the cosine interaction is another important characteristic of the model. It is obviously relevant for large $\varphi_0$ values of the initial phase. Furthermore, it plays an important role in smoothing the Josephson oscillations for general values of the other parameters.

{We remark that our simulations can be considered reliable only up to the onset of revivals, at approximately ($70\,\textrm{ms}$). This timescale corresponds to the propagation time required for correlations originating near the edges to reach the centre of the condensate. Since the boundary regions are treated only within our simplified bulk description, the subsequent revival dynamics is expected to be sensitive to details that are not captured reliably by the present model. Importantly, however, the expectation value ($\langle\cos\varphi\rangle$) is already stabilised close to unity before the onset of the revivals. Our conclusions regarding the mechanism underlying phase locking are therefore unaffected by this limitation. For the same reason, phase locking itself cannot be attributed to a boundary effect, as it develops well before the central region becomes correlated with the edges.}

{The influence of the boundary regions becomes apparent when comparing cases 111 and 110. When both the nonlinearity and the inhomogeneity are included, the common-mode coupling visibly suppresses the revivals. Its effect is negligible prior to their onset, consistent with the validity of the large-density approximation over nearly the entire condensate, but becomes significant in the narrow regions near the edges, where the density vanishes, and the common-mode coupling becomes the dominant interaction term. Consequently, its influence is primarily evident in the dynamics of revival and is attributable to boundary effects. This provides a useful interpretation of the difference between cases 111 and 110 (c.f.~Fig.~\ref{fig:8fold}), but at the same time emphasises that the revival dynamics is strongly sensitive to the treatment of the edges and should therefore not be regarded as a quantitatively reliable prediction of the present model.}

\subsection{Further refinements}

\begin{figure*}
    \centering
    \includegraphics{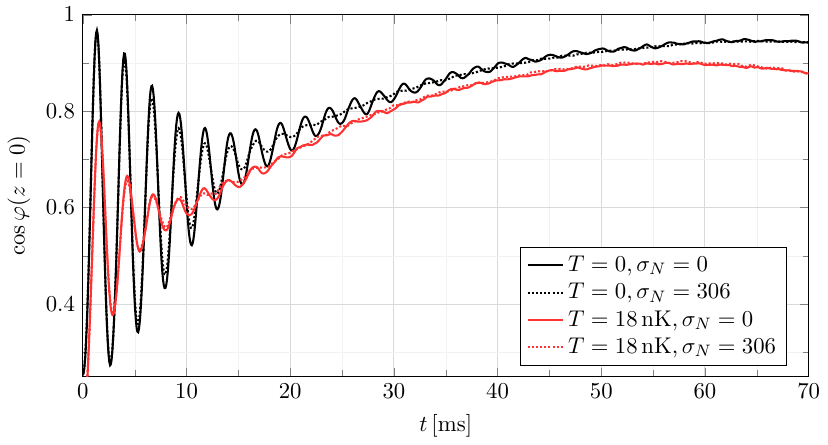}
    \caption{Analysing the effect of finite temperature and the fluctuation of the initial number of atoms considering the $\cos\varphi^r$ expectation value. Parameters: $N = 3300, J/(2\pi\hbar) = 19.0\,\text{Hz}, \varphi_0^r = -1.3,  \sigma_{\varphi_0^r} = 0.2,$.}
    \label{fig:TvsN}
\end{figure*}

To make our analysis more realistic, we considered several further refinements, including finite temperature and uncertainty in the initial particle number loaded into the system at the start of each experimental run.

\subsubsection{Finite temperature}\label{subsec:finite-temp}

We considered runs at finite temperature and found that it leads to an additional damping in the relaxation of $\langle\varphi^r(t)\rangle \rightarrow 0$. Additionally, it also decreases the asymptotic value of the phase $\langle\cos\varphi^r(t)\rangle$ (c.f.~Fig.~\ref{fig:TvsN}). Since the experimentally observed phase locking was quite robust, we expect that the temperature of the relative degrees of freedom was considerably smaller than the energy scales governing phase locking. Therefore, in order to keep things simple, from now on, we consider simulations at zero temperature (if not indicated otherwise).

\subsubsection{Particle number variance}

Another refinement is to account for the uncertainty in the initial parameters. Here, we limit ourselves to the uncertainty of the initial number of particles $N$. In the experiment, the data points are calculated as the average of multiple single runs; therefore, fluctuations in the initial parameters result in additional damping due to decoherence.

In contrast to thermal fluctuation, the decoherence resulting from the variance of the initial particle number leads to damping only in $\langle\varphi^r(t)\rangle$. The uncertainty of the initial parameters does not change the expected variance of $\varphi^r$ and therefore does not dampen the expectation value of $\langle\cos\varphi^r\rangle$.

The simplest way to characterise the distribution of the atom number $N$ is given by a Gaussian distribution. However, it is not clear how the standard deviation $\sigma_N$ is related to the total number of particles. Moreover, one may consider a post-selection approach (neglecting runs with too large a particle deviation, cf.~\cite{2021PhRvR...3b3197M}); however, a sharp cut-off is experimentally unrealistic and yields an unphysical $\approx \text{sinc}(\tilde\omega t)$ modulation. Based on these considerations, we adopt a simple Gaussian distribution, assuming the standard deviation is well-defined by the corresponding experimental setup.

\section{The full dynamics in the experimental parameter regime}

\label{sec:fulldyn}

\subsection{Simulations}

We also ran a full simulation in the parameter regime relevant for the experiment \cite{2018PhRvL.120q3601P}. The results are shown in Fig. \ref{fig:comparism}. From the experimental settings, we fixed the following values for the parameters 
\begin{align}
    \begin{aligned}
    &N = 3300\,,\quad \sigma_N = 300\,,\quad 
    \\
    &\omega_{||} = 2\pi\cdot 12\,\text{s}^{-1}\,,\quad
     J/(2\pi\hbar) = 19\,\text{Hz}\,,
     \\
     &\varphi_0^r = -1.3\,,\quad  \sigma_{\varphi_0^r} = 0.2,
     \end{aligned}
\end{align}
for the average number of atoms and its variance, the longitudinal trap frequency, the Josephson coupling\footnote{The Josephson coupling is determined from the frequency of the oscillations.}, and the initial phase, respectively. The temperature is set to $T=0$ as explained in Subsection \ref{subsec:finite-temp}, and the only variable for which we had a freedom of choice is the initial time offset to match the start of the experimental data taking. 

We found that with a single choice of this time offset, the simulated curves accurately capture the characteristics of the experimental data. Additionally, we also displayed the theoretical evolution of the expectation value of the cosine of the relative phase. The reason is that relaxing the average to zero does not necessarily signify phase locking, which also requires the variance to approach zero. The evidence for the latter is that the cosine of the phase relaxes to a value close to $1$.

\subsection{Timescales}\label{subsec:timescales}

The behaviour of the system can be understood on three different hierarchical timescales.

The first is the timescale of the Josephson oscillations $1/f_J$, for which the dynamics of the relative mode is well described by the sine-Gordon model in a homogeneous potential. We observe no qualitative differences (other than a renormalisation of the Josephson frequency) compared to the inhomogeneous model and the model including the common-mode coupling (see Fig.~\ref{fig:8fold}).

The next timescale is the relaxation time corresponding to $\langle\cos\varphi\rangle(t)\rightarrow\approx1$. Since phase locking is the result of the trap, the time until the system reaches phase locking is characterised by $\omega_{||}$. As is apparent from Fig. \ref{fig:comparism}, this timescale is approximately the same as the relaxation time of the oscillations in $\langle\varphi\rangle(t)\rightarrow0$. We investigate the dependence of the relaxation time on experimental parameters in the next section.

The sine--Gordon model is well-known to be integrable on a homogeneous background. At the same time, it does not show a full relaxation to a phase-locked state; rather, it only has a (universal) rephasing dynamics, for which the expectation value of the relative phase approaches a non-zero value \cite{2013PhRvL.110i0404D,Horvath2019}. The inhomogeneity induced by the trap breaks integrability, and so one naturally expects a profound change in the relaxation dynamics. However, the detailed connection between the breaking of integrability and phase locking remains unclear at this stage. Furthermore, we observe that the coupling to the common mode does not play a significant role at this timescale, although it also breaks integrability.

The third timescale is the one of the revivals that appear, which is when the coupling to the common-mode can become relevant by further suppressing revivals in the inhomogeneous sine-Gordon model. This feature can be understood by noting that the coupling to the common mode is suppressed by the particle density, which is high in the middle region of the condensate but approaches zero near the ends. Therefore, the common-mode coupling has a negligible effect on the dynamics in the middle region of the condensate for short times, but it can have a large effect on the reflection from the boundaries and, therefore, the revivals. However, because the density approaches zero at the boundaries, our field-theoretic description is only valid in its neighbourhood; therefore, a more detailed and realistic description of this regime requires further research.

\section{Simplified models}\label{sec:comparison}

In this section, we analyse the scaling of the relaxation time $\tau$ and the Josephson frequency $\omega_J$ of $\varphi^r(t, z =0)$ as a function of the input parameters of the model.

\subsection{Fitting functions}

To define $\tau$ and $\omega_J$, we fitted $\varphi(t) \equiv \varphi^r(t,z=0)$\footnote{To avoid edge effects, we considered the local observable in the centre and time evolution before revivals reach the middle point.} with a damped oscillation
\begin{align} \label{eq:fitfn}
    \varphi_\textrm{fit}(t) = \varphi_\textrm{fit}^0 e^{-t/\tau} \sin(\omega_J t + \vartheta_0)\,,
\end{align}
which is the exact analytic form of the leading order behaviour of small quenches (with one-particle contributions) in massive quantum field theories  \cite{Rakovszky_2016}. The sinusoidal form can be obtained from a leading order form factor expansion on the pre-quench basis \cite{2014JPhA...47N2001D,2017JPhA...50h4004D}, while a resummation of higher order terms (in an expansion using the post-quench basis) elucidates the origin of the exponential damping \cite{2014JSMTE..10..035B,2017JSMTE..10.3106C}.

We remark that the start of the quench does not provide a good fit even with non-linear model functions (e.g.~for the two-mode model used in Ref.~\cite{2018PhRvL.120q3601P}). We found robust fitting only by considering long-lived excitations. As the amplitude of the long-living excitations becomes relatively small, we can use the fit function in Eq.~\eqref{eq:fitfn} coming from a leading order small density approximation.

\begin{figure}
    \centering
    \includegraphics[width=1.0\linewidth]{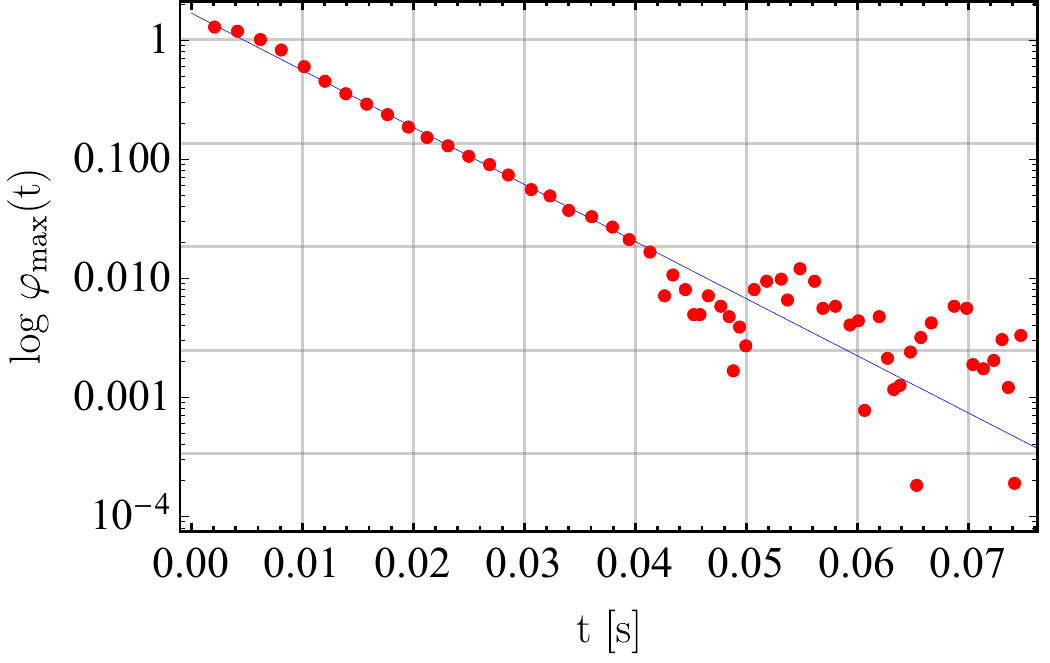}
    \caption{Local extrema of $\varphi(t)$ on the semi-log axis (red dots). The blue line indicates the plateau corresponding to the long-living excitations. Parameters: $T = 0\,\text{nK}, \varphi_0^r = -1.3, \sigma_{\varphi_0^r} = 0.2, \sigma_N = 300, N = 3300, J/(2\pi\hbar) = 31.2\,\text{Hz}$.}
    \label{fig:plateau}
\end{figure}

\begin{figure}
    \centering
    \includegraphics[width=1.0\linewidth]{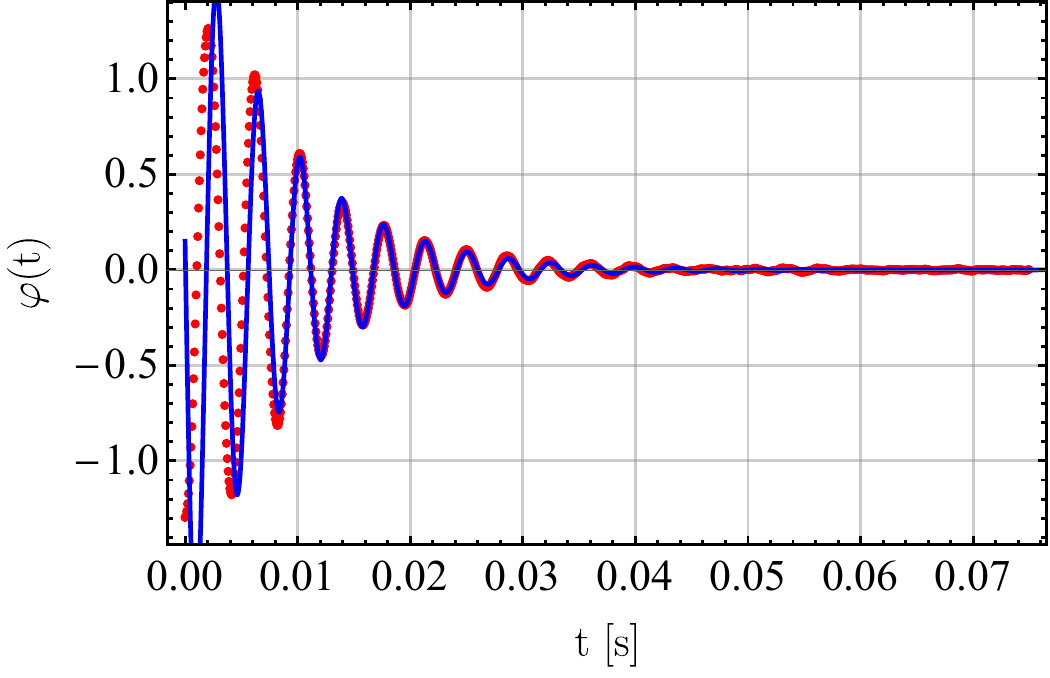}
    \caption{The $\varphi(t)$ expectation value from the TWA simulation (red dots) and the fitted function Eq.~\eqref{eq:fitfn} (blue line). Parameters: $T = 0\,\text{nK}, \varphi_0^r = -1.3, \sigma_{\varphi_0^r} = 0.2, \sigma_N = 300, N = 3300, J/(2\pi\hbar) = 31.2\,\text{Hz}$, fit results: $\varphi_\textrm{fit}^0=2.09, \omega_J = 1694\,\textrm{s}^{-1}, \tau = 8.10\,\textrm{ms}$.}
    \label{fig:fiteg}
\end{figure}

We first smoothed the result of the TWA simulation, $\varphi(t)$, and identified its zeros, local minima, and maxima. Observing the absolute values of the local extremal points on the semi-log plot, we can identify a decreasing plateau corresponding to the relaxation of long-living oscillations (c.f.~Fig.~\ref{fig:plateau}). To capture this regime, we fit Eq.~\eqref{eq:fitfn} between the 3rd and 10th periods of Josephson oscillations. A typical fit is illustrated in Fig.~\ref{fig:fiteg}.

We also briefly analysed the conjugate variable $\Pi(t,z=0)$, which corresponds to the local particle imbalance in the centre. The total particle number difference between the two condensates is extrapolated from this quantity as
\begin{align}
    n := \frac{N_L - N_R}{N_L + N_R} &\approx \frac{4}{3}\frac{N_s}{N}\frac{\Pi_r(0)}{\beta_r}\,,
\end{align}
which is obtained by assuming $\delta\rho^r(z)/\rho_0(z) \approx \delta\rho^r(0)/\rho_0(0)$. This approximation is not necessarily valid in every regime; nevertheless, it provides a way to calculate an effective Josephson coupling ($J_\textrm{fit}$, see Eq.~\eqref{eq:Jfit}) which turns out to be a consistent proxy in the experimental domain of parameters.

\begin{figure}
    \centering
    \includegraphics[width=1.0\linewidth]{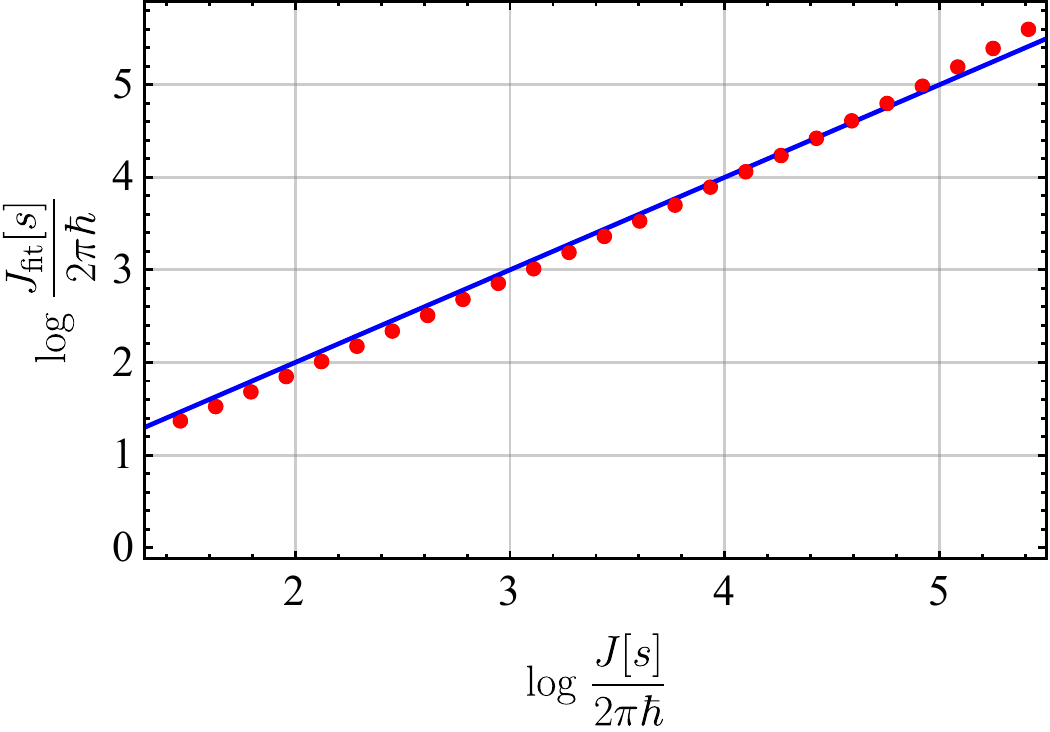}
    \caption{The fitted Josephson coupling $J_\textrm{fit}$ as a function of the input coupling of the simulation $J$ (red dots) compared to the identity function (blue line). Parameters: $N = 3300, T = 0\,\text{nK}, \varphi_0^r = -1.3, \sigma_{\varphi_0^r} = 0.2, \sigma_N = 300$.}
    \label{fig:JvsJ}
\end{figure}

Assuming $\tau \gg 1/\omega_J$, the particle imbalance becomes
\begin{align}
    n_\textrm{fit}(t) = n_\textrm{fit}^0 e^{-t/\tau} \cos(\omega_J t + \vartheta_0)\,.
\end{align}
Considering $\{\varphi(t,0),\Pi(t,0)\}$ as a single-mode oscillator, the effective Josephson coupling can be expressed in terms of the fitted parameters
\begin{align}\label{eq:Jfit}
    J_\textrm{fit} = \frac{\hbar \omega_J}{2} \frac{n_\textrm{fit}^0}{\varphi_\textrm{fit}^0}\,.
\end{align}
We found that the effective Josephson coupling ($J_\textrm{fit}$) is in notable agreement with the raw input parameter of the simulation $J$ (see Fig.~\ref{fig:JvsJ}), which confirms the self-consistency of our approach.

\subsection{Josephson frequency}

\begin{figure*}
    \centering
    \begin{subfigure}{0.48\textwidth}
    \centering
    \caption{}
    \includegraphics[width=1.\linewidth]{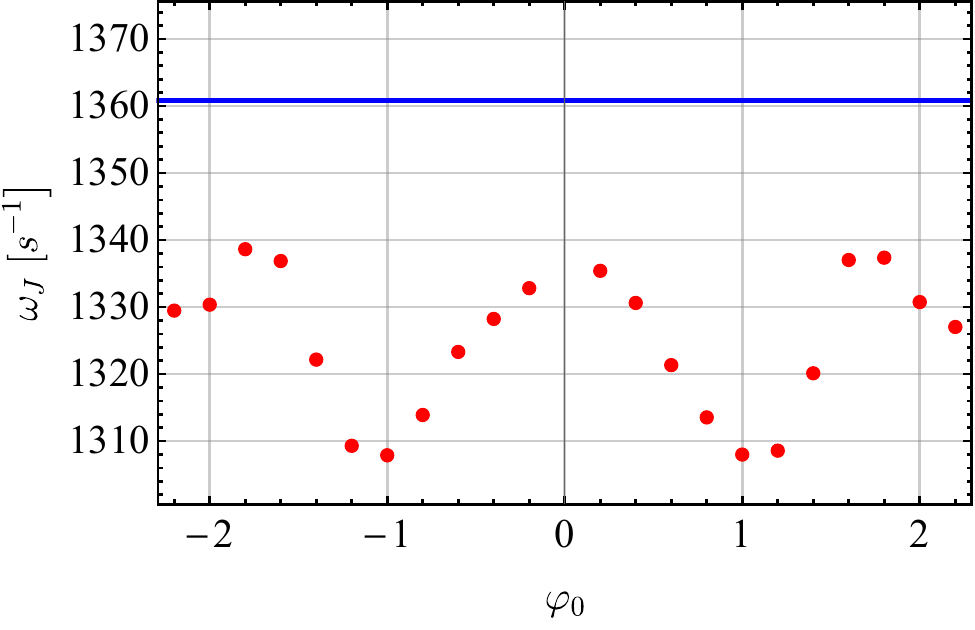}
    \end{subfigure}
    \begin{subfigure}{0.48\textwidth}
    \centering
    \caption{}
    \includegraphics[width=1.\linewidth]{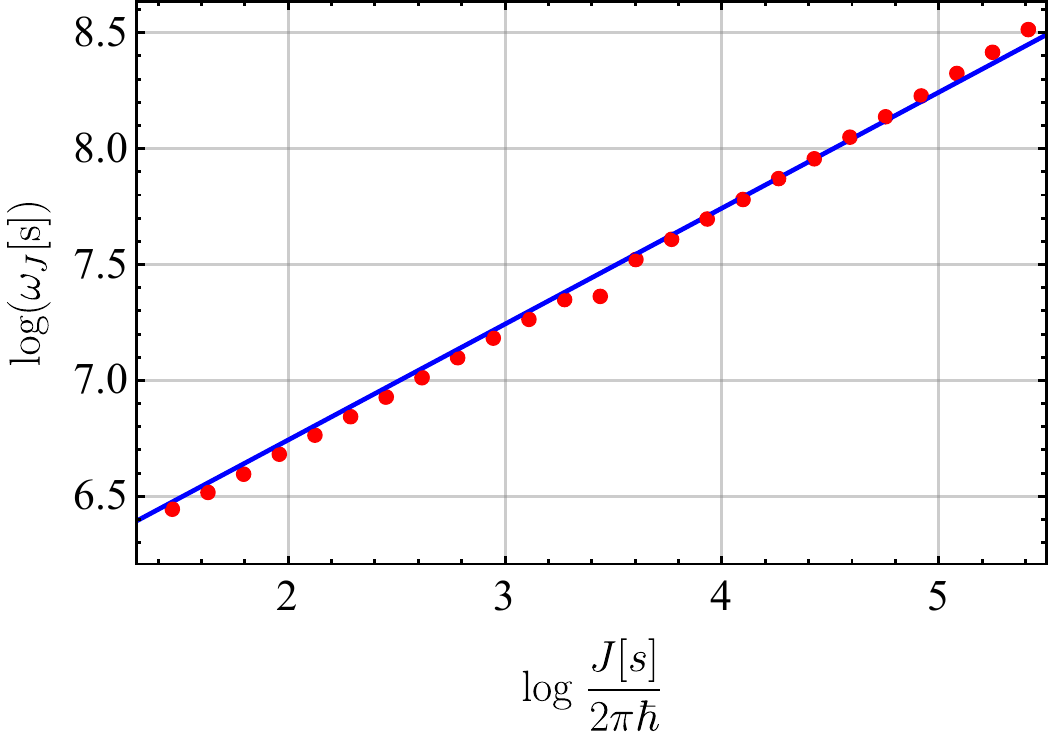}
    \end{subfigure}
    \begin{subfigure}{0.48\textwidth}
    \centering
    \caption{}
    \includegraphics[width=1.\linewidth]{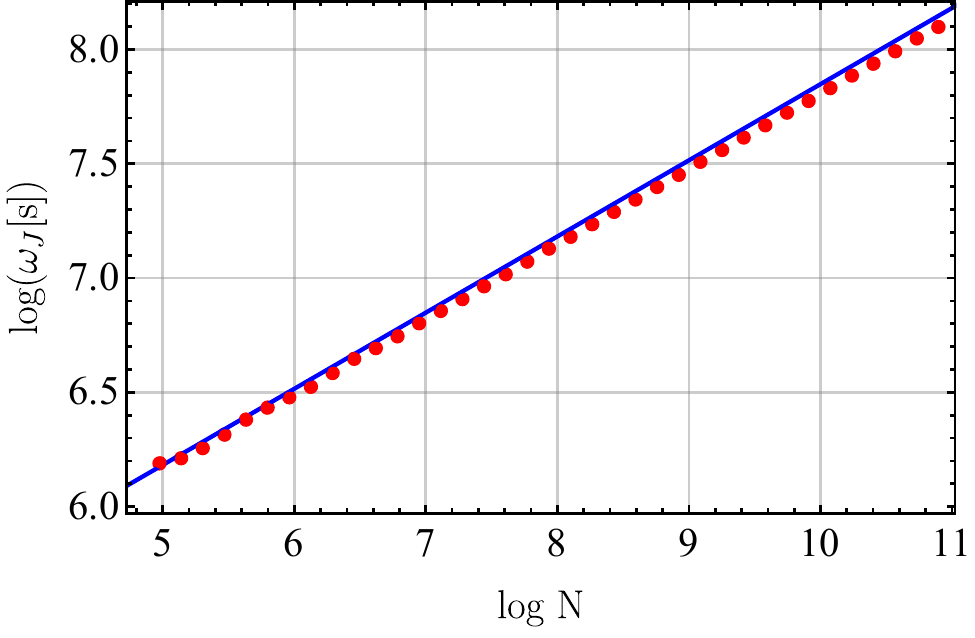}
    \end{subfigure}
    \begin{subfigure}{0.48\textwidth}
    \centering
    \caption{}
    \includegraphics[width=1.\linewidth]{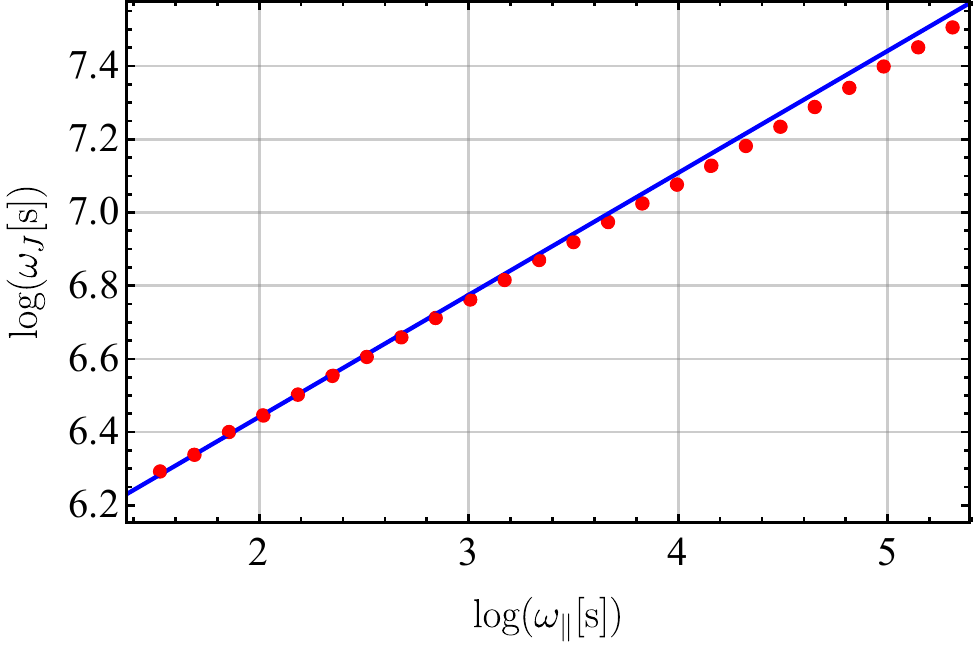}
    \end{subfigure}
    \begin{subfigure}{0.48\textwidth}
    \centering
    \caption{}
    \includegraphics[width=1.\linewidth]{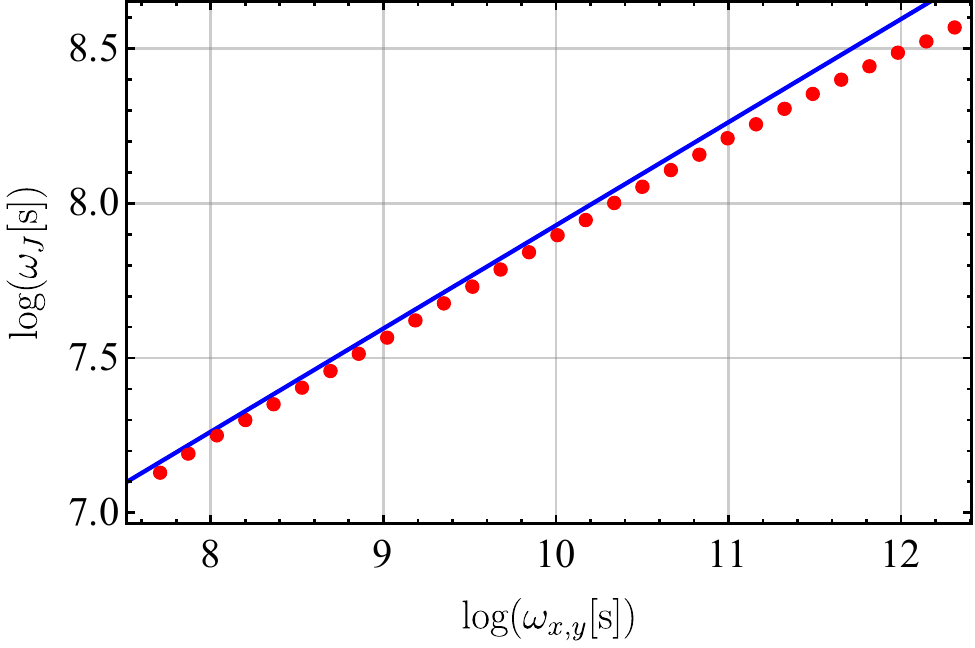}
    \end{subfigure}
    \caption{The Josephson frequency as a function of \textbf{(a)} the initial phase difference $\varphi_0$, \textbf{(b)} the Josephson coupling $J$, \textbf{(c)} the number of atoms $N$, and the strength of \textbf{(d)} the longitudinal and \textbf{(e)} the transversal trapping potential, $\omega_{||}$ and $\omega_{\perp}$. Blue lines indicate the theoretical prediction of the linearised homogeneous model. Parameters: $T = 0\,\text{nK}, \varphi_0 = -1.3(2), N = 3300, \sigma_N = 300, J/(2\pi\hbar) = 19\,\text{Hz}$, \textbf{(e)}: $J/(2\pi\hbar) = 30\,\text{Hz}$.}
    \label{fig:scalingoJ}
\end{figure*}

Next, we analysed the Josephson frequency $\omega_J$. It is natural to compare this to the simplest quantum field-theoretic approximation for the leading-order behaviour. For small quenches, the middle region of the condensate can be approximated with the free (homogeneous) massive boson. The parameters of this free boson are chosen such that the local parameters (particle density $n$, sound velocity $c$, Luttinger parameter $K_0$) match the corresponding ones at the midpoint $z=0$. The initial state is modelled as a coherent state corresponding to an initial expectation value $\varphi_0$. It is then straightforward to express the oscillation frequency with the parameters of the condensate as
\begin{align} \label{Eq:oJ}
    \omega_J = \sqrt{\frac{8\pi^2 \tilde J c_0 n_0}{K_0}} = 2\pi \left(\frac{3^2\tilde J^3 N^2g_\mathrm{1D}^2\omega_{||}^2 m}{2^4\pi^3\hbar^3}\right)^{1/6}\,,
\end{align}
where $\tilde J = J / (2\pi\hbar)$. We check this analytic result against the simulation results obtained in our model.\footnote{The Josephson frequency is obtained by fitting the aforementioned model to the simulated data.} Fig.~\ref{fig:scalingoJ} shows the frequency of the Josephson oscillations as a function of the Josephson coupling, the number of atoms, and the strengths of the trapping potential, compared to the theoretical prediction \eqref{Eq:oJ}. We found good agreement between the theoretical prediction and the simulated results. 

The agreement between the theoretical prediction of the linearised model and the simulation proved robust even for large initial phase differences, indicating that the dynamics is dominated by low-energy sine-Gordon breather quasi-particles, which are well approximated within the linearised model, after an initial transient that is omitted from the fitting.

\subsection{Relaxation time}

\begin{figure*}
    \centering
    \begin{subfigure}{0.48\textwidth}
    \centering
    \caption{}
    \includegraphics[width=1.\linewidth]{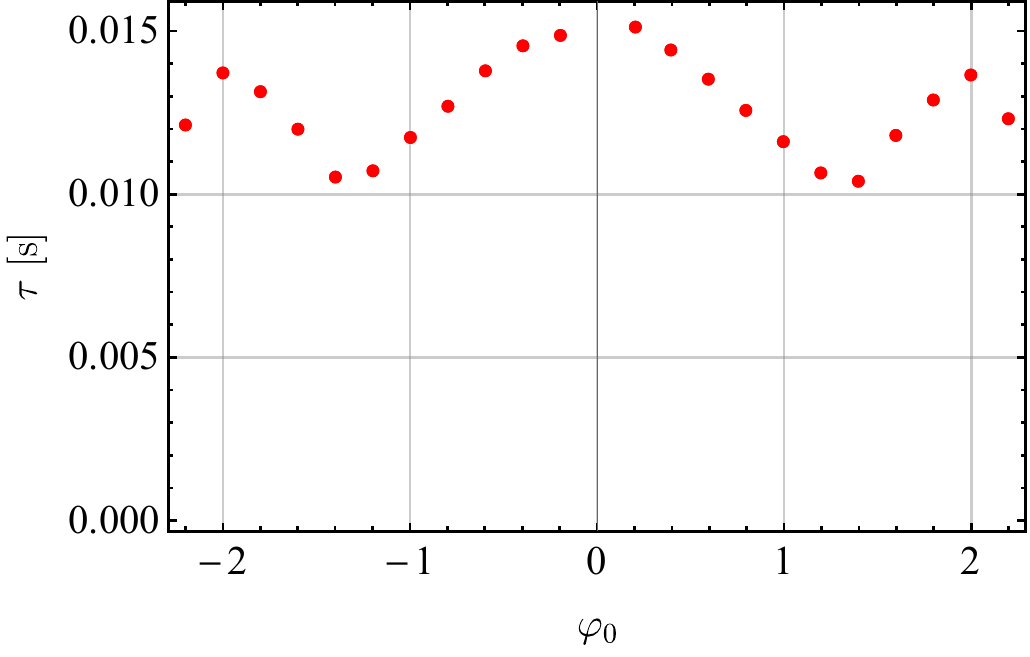}
    \end{subfigure}
    \begin{subfigure}{0.48\textwidth}
    \centering
    \caption{}
    \includegraphics[width=1.\linewidth]{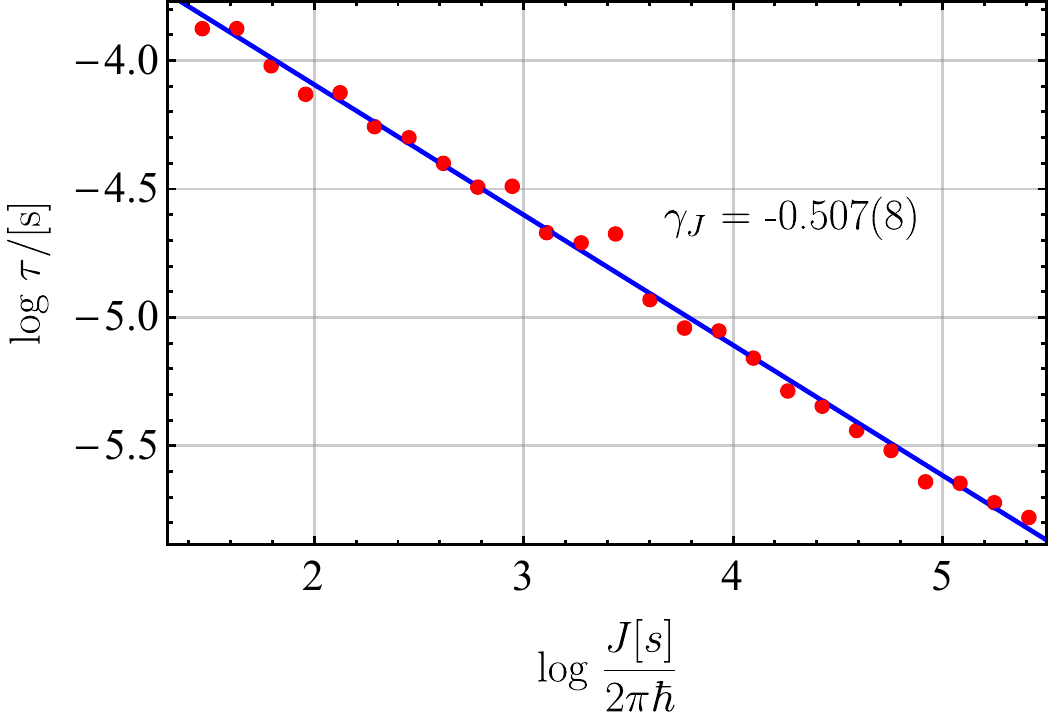}    
    \end{subfigure}
    \begin{subfigure}{0.48\textwidth}
    \centering
    \caption{}
    \includegraphics[width=1.\linewidth]{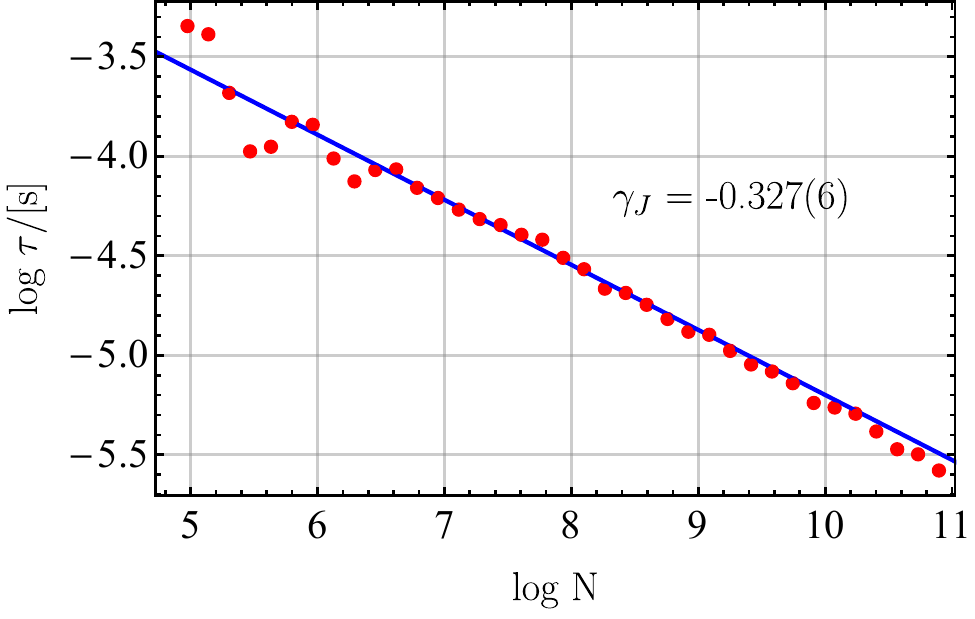}    
    \end{subfigure}
    \begin{subfigure}{0.48\textwidth}
    \centering
    \caption{}
    \includegraphics[width=1.\linewidth]{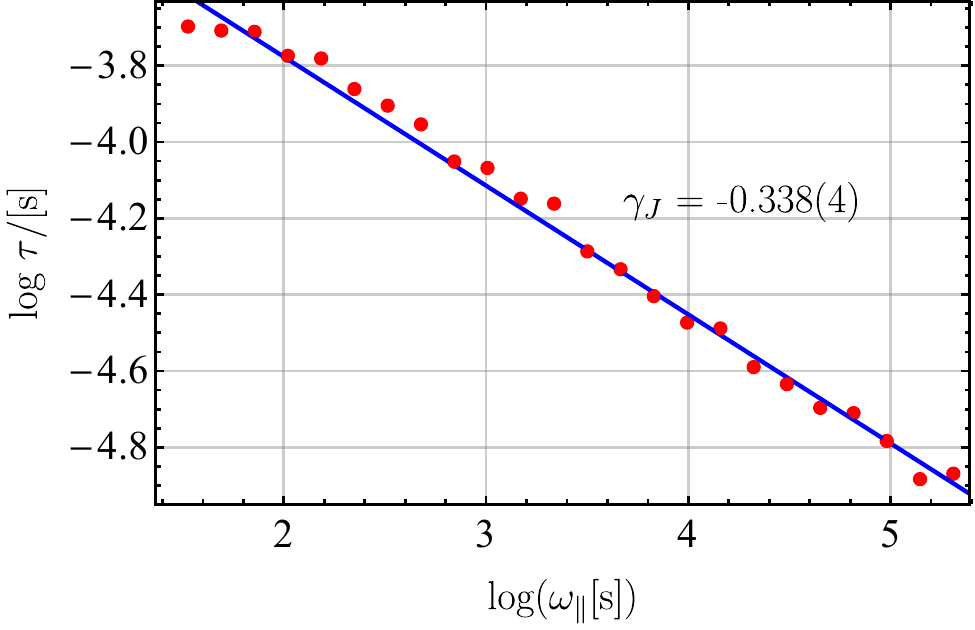}
    \end{subfigure}
    \begin{subfigure}{0.48\textwidth}
    \centering
    \caption{}
    \includegraphics[width=1.\linewidth]{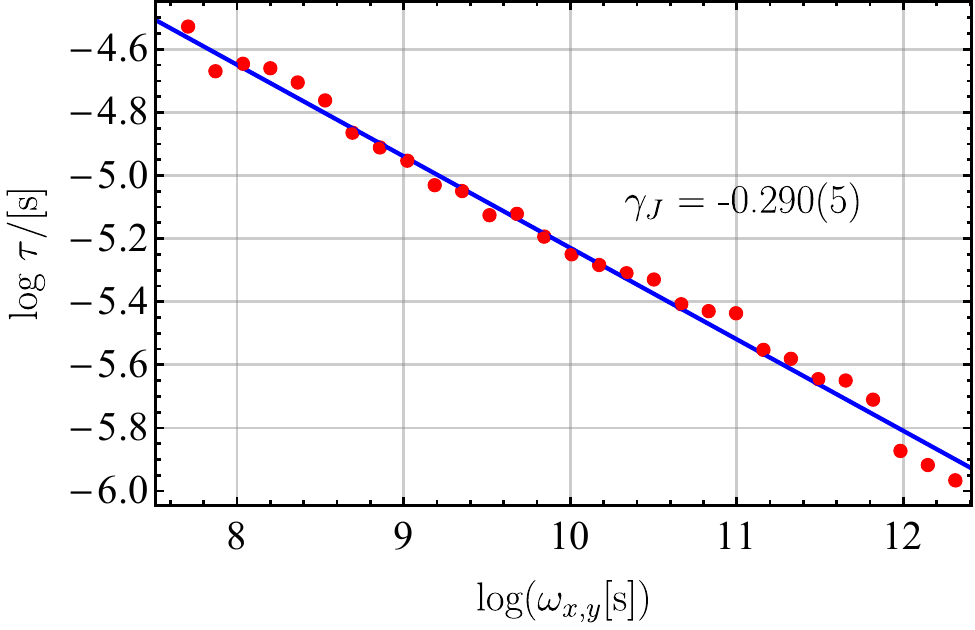}
    \end{subfigure}
    \caption{The relaxation time $\tau$ as a function of \textbf{(a)} the initial phase difference $\varphi_0$, \textbf{(b)} the Josephson coupling $J$, \textbf{(c)} the number of atoms $N$, and the strength of \textbf{(d)} the longitudinal and \textbf{(e)} the transversal trapping potential, $\omega_{||}$ and $\omega_{\perp}$. We also showed the fitted line (blue) and the corresponding exponent $\gamma$. Parameters: $T = 0\,\text{nK}, \varphi_0 = -1.3(2), \sigma_N = 300$, $N = 3300, J/(2\pi\hbar) = 19.0\,\text{Hz}$.}
    \label{fig:scalingtau}
\end{figure*}

Next, we considered the damping time $\tau$ at different parameter settings (see Fig.~\ref{fig:scalingtau}). We found that the damping time only shows a slight oscillatory dependence on the initial global phase $\varphi_0$. On the other hand, we found a decreasing tendency of $\tau$ as a function of the Josephson coupling $J$, the total number of particles $N$ and the strengths of the trapping potential. Assuming a power-law dependence, we fitted a line to the logarithmic plot to extract the corresponding exponents. We found that the relaxation time scaled with the power $-1/2$ of the Josephson coupling $J$, and the power $-1/3$ of the particle number $N$ and longitudinal trap strength $\omega_{||}$. For the case of the transversal potential strength, $\omega_\perp$, the exponent is not absolutely clear (somewhere between $1/4$ and $1/3$); a wider logarithmic range should be covered to give a more accurate result. Unfortunately, the parameters cannot be changed indefinitely, as one must observe a reasonable number of Josephson oscillations before revivals return from the edges.

Our results for the scaling of $\tau$ as a function of the parameters $J$ and $N$ deviate from the results given in previous studies \cite{2018PhRvL.120q3601P,2021PhRvR...3b3197M}. In this respect, we emphasise that our simulations are based on an effectively one-dimensional model that does not capture all 3D phenomena. One such correction is the local swelling of the condensate as the linear particle density increases, which results in position-dependent tunnel coupling. Taking these effects into account requires a substantial extension of our model, perhaps along the lines of Ref.~\cite{2021ScPP...10...90V}; we leave these considerations for future studies. We also note that the three-dimensional stochastic Gross-Pitaevskii model considered in Ref.~\cite{2021PhRvR...3b3197M} could successfully capture the experimentally observed dependence of the relaxation time on the input parameters.

Another source of the discrepancy may be the definition of the relaxation time. While the damped oscillation in Eq.~\eqref{eq:fitfn} is the low-amplitude approximation of the two-mode model result in Ref.~\cite{2018PhRvL.120q3601P}, the fitted regime in time is different: instead of the long-living excitations, the experiment focuses on a shorter timescale where the amplitude is still large.

A further question is the proper form of the initial state in the effectively one-dimensional model. For simplicity, following \cite{Horvath2019,2024PhRvB.109a4308S} we considered the non-zero mode part of the initial state as a zero (or finite) temperature equilibrium state, while the zero mode realised the initial angle $\varphi_0$, which is a coherent state. Considering more realistic initial states that take into account the effect of the preparation procedure by splitting the condensate \cite{2015PhRvA..91b3627F,2017EPJST.226.2763F} is left for further study.

Additionally, as discussed at the end of Sec. \ref{sec:fulldyn}, near the edges of the harmonic longitudinal trap, where the density approaches zero, the field-theoretic approximation is not valid and the Luttinger liquid description breaks down. In our analysis, the Hamiltonian \eqref{eq:H} used for the simulations is regular (finite) even at the edges; however, the next order interaction (in particle density) would blow up. To avoid unreliable results, we analysed quantities located at the centre and only considered time evolution before revivals appear. Developing a robust method to model the edges requires further analysis.

\section{Conclusions}\label{sec:conclusions}

We considered an effective one-dimensional quantum field theory model of the coupled bosonic condensate simulator of quantum sine-Gordon field theory. Previously, experiments established it as a good simulator in equilibrium \cite{2017Natur.545..323S,2018PhRvA..98b3613B}; however, its non-equilibrium dynamics displayed phase-locking behaviour \cite{2018PhRvL.120q3601P} in contrast to the universal rephasing predicted from theory \cite{2013PhRvL.110i0404D,2015PhRvA..91b3627F,2017EPJST.226.2763F,Horvath2019}. 

\begin{figure}
    \centering
    \includegraphics{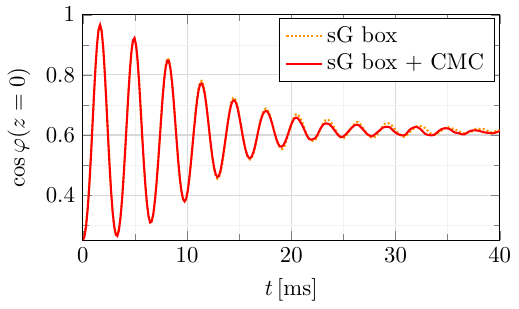}
    \caption{Phase relaxation of the pure sine--Gordon model compared to the sine--Gordon dynamics with the common mode coupling (CMC) in a flat-bottom (box) longitudinal trapping potential. (Data is the same as in Fig.~\ref{fig:8fold}.) Parameters: $N = 3300, J/(2\pi\hbar) = 19.0\,\text{Hz}, T = 0.0\,\text{nK}, \varphi_0^r = -1.3, \sigma_{\varphi_0^r} = 0.2, \sigma_N = 0$.}
    \label{fig:concl}
\end{figure}

We have demonstrated that the phase-locking dynamics can be accounted for by considering the effect of the longitudinal trapping potential. The role of the longitudinal trapping was already found in the stochastic Gross-Pitaevski approach \cite{2021PhRvR...3b3197M}. However, our approach reaches this conclusion in a one-dimensional field-theory setting, suggesting that the experiment can be considered a good simulator of sine-Gordon field theory on an inhomogeneous background. We also found that the coupling between the relative phase (corresponding to the sine-Gordon field) and the average phase (common mode) does not play a significant role during the time interval over which boundary effects can be neglected. As a consequence, we fully expect that a box trap prepared by shaping the trapping potential, as proposed in Ref. \cite{2019OExpr..2733474T}, can achieve an accurate simulation of sine-Gordon dynamics out of equilibrium as illustrated in Fig.~\ref{fig:concl}. {At least, within our framework.}

{While the present inhomogeneous one-dimensional field-theory description successfully accounts for phase locking, it does not reproduce the experimentally observed scaling of the relaxation time with system parameters reported in Ref. \cite{2018PhRvL.120q3601P}, in contrast to the stochastic Gross-Pitaevskii approach of Ref. \cite{2021PhRvR...3b3197M} which does reproduce this scaling. We conjecture that the discrepancy arises from physical effects not captured by our effective one-dimensional description, most notably three-dimensional corrections and the treatment of low-density edge regions. Establishing this connection quantitatively remains an important problem for future work.}

{We also note that the treatment of the low-density edge region and the associated boundary conditions remains a subtle issue. In particular, Ref. \cite{2024PhRvB.109c5118B} emphasised that the treatment of the Luttinger parameter near the edges of the trap may have important consequences for the resulting boundary conditions and the propagation of sine-Gordon quasiparticles. Such effects are expected to primarily influence the long-term revival dynamics observed in our simulations. However, the fact that phase locking is already established before boundary-induced correlations can propagate from the edges to the centre of the condensate suggests that our identification of trap-induced inhomogeneity as the mechanism underlying phase locking is robust against modifications of the boundary description. A more detailed investigation of the interplay between boundary physics and the long-time dynamics remains an interesting direction for future work.}

We close by mentioning a few interesting open questions. One of them is to take into account deviations from the strictly one-dimensional behaviour, i.e., 3D effects, which are thought to account for the anomalous scaling of the relaxation time and the Josephson frequency with the number of atoms in the trap. The other is modelling the behaviour near the edges. In the case of the parabolic longitudinal trap, the density of the condensate approaches zero at the edge; therefore, the large density approximation ceases to be valid. For a box trap, we expect that the sine-Gordon field satisfies Neumann boundary conditions at the edge, which is an integrable boundary condition with exactly known reflection factors \cite{1994IJMPA...9.3841G} and boundary excitation spectrum \cite{2001NuPhB.614..405B}, which makes it an interesting question whether these predictions can be verified experimentally.  

\subsection*{Author contributions}
G.T. conceived the work, selected and outlined the methods, guided their implementation and verification, supervised the work and wrote the manuscript. B.F. implemented the simulations, performed the analysis, contributed to the interpretation of the results, development of the ideas and writing the manuscript.

\acknowledgements{
We are grateful to J. Schmiedmayer for inspiring discussions and for providing access to some of their unpublished experimental results. This work was supported by the HUN-REN Hungarian Research Network through the Supported Research Groups Programme, HUN-REN-BME-BCE Quantum Technology Research Group (TKCS-2024/34), and also by the National Research, Development and Innovation Office of Hungary (NKFIH) through OTKA Grant No. ANN 142584. The NKFIH also provided partial support to G.T. via the Grant “Quantum Information National Laboratory“ with Grant No. 2022-2.1.1-NL-2022-00004. B.F. was also partially supported by the Doctoral Excellence Fellowship Programme (DCEP), funded by the National Research Development and Innovation Fund of the Ministry of Culture and Innovation and the Budapest University of Technology and Economics. 
}
\bibliographystyle{utphys}
\bibliography{phase_locking_refs}

\clearpage

\appendix

\section{Summary of model parameters}\label{app:parameters}

In this appendix, we summarise all the parameters of our model. It is important to precisely separate the input parameters that vary independently and to determine all the physical characteristics of the system.

Input parameters of the experiment (which are directly given or can be tuned):
\begin{itemize}
    \item atomic mass $m$ (for $^{87}\text{Rb}$, $m = 86.90918 \cdot 1.660539\cdot 10^{-27} = 1.44316\cdot10^{-25}\,\text{kg}$),
    \item one-dimensional interaction strength $g_\mathrm{1D} \approx 2\hbar \omega_\perp a_s$ (for $^{87}\text{Rb}$, $a_s= 5.24\,\text{nm}$, and for the experimental setup considered, $\omega_\perp = 2\pi \cdot 1500\,\text{Hz}$),
    \item total number of atoms $N$ (typically between 750 and 4500) and its uncertainty$\sigma_N$ (if its known),
    \item longitudinal trapping potential $\omega_{||}$ (for the experimental setup considered, $\omega_\perp = 2\pi \cdot 12\,\text{Hz}$ if not indicated otherwise),
    \item Josephson coupling $\tilde J = J/(2\pi\hbar)$,
    \item the initial total relative phase $\varphi_0^r$ and its standard deviation $\sigma_{\varphi_0^r}$,
    \item temperature $T$ (the temperature of the experimental condensate is $18\,\text{nK}$ before the splitting procedure; however, as argued in Section \ref{subsec:finite-temp}, the relative degrees must be substantially colder, therefore we chose the temperature as $0$).
\end{itemize}
These input parameters determine the physical system, and they can be changed independently. From these, we can calculate many other physical quantities relevant to our study, such as
\begin{itemize}
    \item length of the condensate $L$,
    \item 1D density at the midpoint ($z=0$):  $n_0$,
    \item sound velocity at the midpoint: $c_0$,
    \item Luttinger parameter at the midpoint: $K_0$,
    \item chemical potential, $\mu$,
    \item Josephson frequency, $\omega_J$.
\end{itemize}
The relations determining these parameters are given by
\begin{align}
    L &= \left(\frac{6g_\mathrm{1D}N}{m\omega_{||}^2}\right)^{1/3}\,,\nonumber\\
    n_0 &= \frac{3^{2/3}}{2^{7/3}}\left(\frac{m\omega_{||}^2N^2}{g_\mathrm{1D}}\right)^{1/3}\,,\nonumber\\
    c_0 &= \frac{3^{1/3}}{2^{7/6}}\left(\frac{g_\mathrm{1D}N\omega_{||}}{m}\right)^{1/3}\,,\nonumber\\
    K_0&=\pi\hbar\frac{3^{1/3}}{2^{7/6}}\left(\frac{N\omega_{||}}{mg_\mathrm{1D}^2}\right)^{1/3}\,,\nonumber\\
    \mu &= \left(\frac{3^{2}}{2^{7}}g_\mathrm{1D}^2N^2\omega_{||}^2m\right)^{1/3}\,,\nonumber\\
    \omega_J &= 2\pi \left(\frac{3^2\tilde J^3 N^2g_\mathrm{1D}^2\omega_{||}^2 m}{2^4\pi^3\hbar^3}\right)^{1/6}\,.\label{eq:parameter_relations}
\end{align}
The following relations are also convenient:
\begin{align}
    R_\text{TF}^2 &= \frac{2\mu}{m\omega_{||}^2}\,,\nonumber\\
    n_0 &= \frac{\mu}{g_\mathrm{1D}}\,,\nonumber\\
    N &= \frac{4}{3}n_0 R_\text{TF}\,,\nonumber\\
    \mu &= \left(\frac{3}{4}g_{1D}N\sqrt{\frac{1}{2}m\omega_{||}^2}\right)^{2/3}\,,\nonumber\\
    m &= \frac{\pi\hbar n_0}{K_s^0 c_0}\,,\nonumber\\
    N_s&= \left\lfloor\frac{3^{2/3}}{2^{5/6}}\left(\frac{mg_\mathrm{1D}^2 N^2}{\omega_{||}\hbar^3}\right)^{1/3}\right\rfloor\,,\nonumber\\ \label{eq:gvsparam}
    g &= 2\pi \tilde J \frac{2^{3/2}}{\omega_{||}}\,\nonumber\\
    g_\mathrm{1D} &= \frac{\pi\hbar c_0}{K_0}\,.
\end{align}
The lattice parameters are chosen according to
\begin{align}
    N_s &= \left\lfloor\frac{Lmc_0}{\hbar}\right\rfloor\,,\nonumber\\
    a &= \frac{L}{N_s}\,,
\end{align}
so the lattice constant satisfies
\begin{align}
    a &\approx \frac{\hbar}{mc_0}\,.
\end{align}

\end{document}